\documentclass{ws-ijmpb}
\begin{document}

\markboth{Peng-Bo Li} { Efficient scheme for entangled states with
trapped atoms in a resonator}

%%%%%%%%%%%%%%%%%%%%% Publisher's Area please ignore %%%%%%%%%%%%%%
\catchline{}{}{}{}{}
%%%%%%%%%%%%%%%%%%%%%%%%%%%%%%%%%%%%%%%%%%%%%%%%%%%%%%%%%%%%%%%%%%%

\title{EFFICIENT SCHEME FOR ENTANGLED STATES WITH TRAPPED ATOMS IN A RESONATOR}

\author{Peng-Bo Li}
\address{MOE Key Laboratory for Nonequilibrium Synthesis and Modulation of Condensed Matter,\\
Department of Applied Physics, Xi'an Jiaotong University, Xi'an
710049, China\\
lipengbo@mail.xjtu.edu.cn}

%\begin{history}
%\received{Day Month Year} \revised{Day Month Year}
%\accepted{Day Month Year}
%\comby{(xxxxxxxxxx)}
%\end{history}
\maketitle
\begin{abstract}
A protocol is proposed to generate atomic entangled states in a
cavity QED system. It utilizes Raman transitions or stimulated Raman
adiabatic passages between two systems to entangle the ground states
of two three-state $\Lambda$-type atoms trapped in a single mode
cavity. It does not need the measurements on cavity field nor atomic
detection and can be implemented in a deterministic fashion. Since
the present protocol is insensitive to both cavity decay and atomic
spontaneous emission, the produced entangled states may have some
interesting applications in quantum information processing.
\end{abstract}


PACS Number(s): 03.67.Mn, 42.50.Pq

\section{Introduction}
Entangled states not only could be utilized to test fundamental
quantum mechanical principles such as Bell's inequalities
\cite{Bell} but also play a central role in practical applications
of quantum information processing \cite{quantum_information}, such
as quantum computation\cite{Shor,Prl-79-325}, quantum
teleportation\cite{Prl-70-1895}, and quantum
cryptography\cite{Prl-67-661}. It is generally believed that atoms
are good candidates for storing quantum information and are natural
quantum information processors. Therefore, producing atomic
entangled states are particularly significant. In the context of
cavity QED \cite{Kimble,Sci298}, numerous proposals have been
presented for generating atomic entangled states
\cite{mplb,Li,Guo,prl-87-037902,Prl-89-237901,Prl-90-127903,Prl-90-217902,Prl-90-253601,Prl-91-067901,prl-92-117902,pra-68-033817,pra-72-012339}.
Though these proposals seem very promising, some rely on
measurements on photons, which impairs their performance; some are
not immune to atomic spontaneous emission.

In this paper, we propose a protocol for the realization of atomic
entangled states with a cavity QED system. It consists of two
three-state $\Lambda$-type atoms and a single mode cavity. We show
that, through suitably choosing the detunings and intensities of
fields, Raman transitions or stimulated Raman adiabatic passages
(STIRAP) \cite{RMP-70-1003} between the two atoms can be achieved,
which can be utilized to produce the atomic entangled states. This
proposal could be implemented in a deterministic fashion, thus
requiring no measurements on cavity field and atomic states. Because
the atomic excited states and cavity mode excitations are not
involved in this process, this protocol is very robust against
atomic spontaneous emission and cavity photon decay. Since the
STIRAP techniques \cite{RMP-70-1003} are utilized, the protocol is
very robust against moderate fluctuations of experimental
parameters. With presently available experimental setups in cavity
QED, this proposal could be implemented.

\section{Generation of atomic entangled states through Raman transitions}

Consider the case of two three-state $\Lambda$-type atoms trapped in
a single mode cavity. As sketched in Fig. 1, each atom has the level
structure of a $\Lambda$ system with two stable ground states $\vert
0\rangle$ and $\vert1\rangle$, and an excited state $\vert
e\rangle$. The classical field of frequency $\omega_L$ drives
dispersively the transition $\vert 0\rangle\leftrightarrow\vert
e\rangle$ with the Rabi frequency $\Omega_i(i=1,2)$ and detuning
$\Delta=\omega_{e1}-\omega_L$. The cavity mode of frequency $\nu$
couples the transition $\vert 1\rangle\leftrightarrow\vert e\rangle$
with the coupling constant $g_i$ and the same detuning
$\Delta=\omega_{e0}-\nu$. For simplicity the coupling constants of
both atoms to the cavity mode are taken to be the same, $g_1=g_2=g$,
but this is not the necessary condition for the analysis. In
addition, we neglect the position dependence of the cavity-atom
coupling strengths by assuming the Lamb-Dicke limit. In the
interaction picture, the associated Hamiltonian under the dipole and
rotating wave approximation is given by (let $\hbar=1$)
\begin{equation}
\hat{H}_I=\sum_{i=1,2}(\Omega_i\hat{\sigma}^i_{e0}e^{i\Delta
t}+g_i\hat{a}\hat{\sigma}^i_{e1}e^{i\Delta t})+\mbox{H.c.},
\end{equation}
where $\hat{\sigma}_{jm}=\vert j\rangle\langle m\vert$ is the atomic
transition operator, and $\hat{a}$ is the annihilation operator for
the cavity mode. We consider dispersive detuning
$|\Delta|\gg|\Omega_i|,|g|$ for each atom. Since level $\vert
e\rangle$ is coupled dispersively with both levels $\vert 0\rangle$
and $\vert 1\rangle$, it can be adiabatically eliminated and atomic
spontaneous emission can be neglected \cite{James}. Then we obtain
the effective Hamiltonian describing the Raman excitations of the
atoms
\begin{eqnarray}
\label{H1}
\hat{H}_{\mbox{eff}}&=&\sum_{i=1,2}(\frac{|g|^2}{\Delta}\hat{a}^{\dag}\hat{a}\hat{\sigma}^i_{11}+
\frac{|g|^2m}{\Delta}\hat{\sigma}^i_{00}\nonumber\\
&&+\frac{|\Omega_ig^*|}{\Delta}\hat{a}^\dag\hat{\sigma}^i_{10}
+\frac{|\Omega_i^* g|}{\Delta}\hat{a}\hat{\sigma}^i_{01}),
\end{eqnarray}
where we chose $\Omega_i$ in phase with $g$. We have included an
energy shift
$\Delta_m=\frac{|g|^2m-|\Omega_i|^2}{\Delta}(m=0,1,...)$ to level
$\vert 0\rangle$ for each atom, which could be implemented through
the action of external classical fields. The number $m$ is
introduced for convenience, which can be determined from the
expression for the energy shift $\Delta_m$, and can be controlled
through tuning the external classical fields. For instance, if we
tune the external classical fields such that the energy shift has
the value $-\frac{|\Omega_i|^2}{\Delta}$, this corresponds to
choosing $m=0$. In the same way, we can choose $m$ for other values.

Assume that atoms 1 and 2 are initially prepared in their stable
ground states $\vert 0\rangle_1$ and $\vert 1\rangle_2$, and the
cavity field is in vacuum state $\vert 0\rangle_C$. Then the
dynamics is confined to the subspace of the collective energy levels
of the two atoms and cavity mode $\{\vert
01;0\rangle,\vert11;1\rangle,\vert 10;0\rangle\}$, where $\vert
ij;k\rangle=\vert i\rangle_1\vert j\rangle_2\vert k\rangle_C
(i,j,k=0,1)$ describes a system with the atoms in state $\vert
i\rangle_1\vert j\rangle_2$ and cavity in Fock state $\vert
k\rangle_C$. Then we can write the Hamiltonian of Eq. (\ref{H1}) in
this subspace as
\begin{eqnarray}
\label{H2} \hat{H}_{\mbox{eff}}&=&\frac{m|g|^2}{\Delta}(\vert
10;0\rangle\langle 0;01\vert+\vert 01;0\rangle\langle 0;10\vert)
+\frac{2|g|^2}{\Delta}\vert 11;1\rangle\langle
1;11\vert\nonumber\\
&&+\frac{|g\Omega_1^*|}{\Delta}\vert 01;0\rangle\langle
1;11\vert+\frac{|g\Omega_2^*|}{\Delta}\vert 10;0\rangle\langle
1;11\vert+\mbox{H.c.},
\end{eqnarray}
which forms a typical $\Lambda$ system. If we assume that $m=0$ and
$\frac{2|g|^2}{\Delta}\gg\{\frac{|g\Omega_1^*|}{\Delta},\frac{|g\Omega_2^*|}{\Delta}\}$,
then we obtain a Raman transition between states $\vert 01;0\rangle$
and $\vert 10;0\rangle$. In this case, through adiabatic elimination
of the state $\vert 11;1\rangle$, we get an effective Hamiltonian
describing the Raman excitation
\begin{eqnarray}
\label{H3} \hat{H}_{\mbox{eff}}&=& \Theta \vert 10;0\rangle\langle
0;10\vert+\mbox{H.c.},
\end{eqnarray}
with $\Theta=\frac{|\Omega_1\Omega_2|}{2\Delta}$ being the Raman
transition rate. The Hamiltonian (\ref{H3}) describes a two photon
Raman transition between two distant atoms trapped in a cavity.

The system is initially prepared in the state $\psi(0)=\vert
01;0\rangle$, then the state evolution of the system is given by
\begin{eqnarray}
\label{E1} \psi(t)=\cos(\Theta t)\vert 01;0\rangle-i\sin(\Theta
t)\vert 10;0\rangle,
\end{eqnarray}
which is an entangled state for the two atoms. If we choose $\Theta
t=\pi/4$, we could obtain the maximally entangled two-atom state
\begin{equation}
\label{E2} \psi_{a}=\frac{1}{\sqrt{2}}(\vert 01\rangle-i\vert
10\rangle),
\end{equation}
which is the well-known EPR state. This entangled state is very
robust because it only involves the ground states of the atoms.

\section{Atomic entanglement through stimulated Raman adiabatic passages}
Although the entanglement mechanism given above could work well for
a pair of atoms, it relies on off-resonance Raman transitions and is
not robust enough. We now extend the idea to the case of
on-resonance STIRAP process between two trapped atoms in a cavity.
We will see that this STIRAP protocol is more robust than the Raman
excitation based scheme. Different from the general STIRAP process
\cite{RMP-70-1003}, the present transfer is between ground states of
two atoms and keeps cavity field from exciting.

Assume that the two trapped atoms are far apart so that they can be
addressed individually by laser beams with time dependent Rabi
frequencies. We also suppose that the amplitudes of
$\Omega_1(t),\Omega_2(t)$ satisfy $\Omega_1=-\xi\Omega_2$ in the
paper, where $\xi$ is a control parameter. If we choose $m=2$, then
from Eq. (\ref{H2}) we can get the following Hamiltonian
\begin{eqnarray}
\label{H4} \hat{H}_{\mbox{eff}}&=&\frac{|g\Omega_1(t)|}{\Delta}\vert
01;0\rangle\langle 1;11\vert-\frac{|g\Omega_2(t)|}{\Delta}\vert
10;0\rangle\langle 1;11\vert+\mbox{H.c.},\nonumber\\
\end{eqnarray}
where we have discarded the constant energy terms. The effective
Hamiltonian (\ref{H4}) describes a typical $\Lambda$ system which is
on resonance. Therefore, dark state exists in this system,
\begin{eqnarray}
\label{DS} \vert\mbox{D}(t)\rangle&=&\cos\theta(t)\vert
01;0\rangle+\sin\theta(t)\vert 10;0\rangle,
\end{eqnarray}
with $\tan\theta(t)=|\Omega_1(t)|/|\Omega_2(t)|$. As a consequence,
adiabatic transfer of population can occur between states $\vert
01;0\rangle$ and $\vert10;0\rangle$ by slowly varying the laser
amplitudes $\Omega_1(t)$ and $\Omega_2(t)$. This procedure resembles
the STIRAP process, which transfers population between ground states
of one atom, but the present transfer is between ground states of
two atoms and keeps cavity field excitation from being involved in
this process. A maximally entangled state can be generated in the
particular case $|\Omega_1(t)|=|\Omega_2(t)|$, i.e.,
$1/\sqrt{2}(\vert 01\rangle+\vert10\rangle)$.  The generated
entangled two-atom state is more robust than previously proposed
entangled atomic states, due to the fact that the atomic excited
states and cavity mode are unpopulated during the process.

To verify the above approximations and STIRAP process, we
numerically simulate the dynamics generated by the full Hamiltonian
(including terms describing cavity decay and atomic spontaneous
emission) and compare it with the results generated by the effective
model (\ref{H4}). In Fig. 2 the numerical results of the system
evolution with decay terms for the atoms ($\Gamma=0.1g$) and cavity
modes ($\kappa=0.1g$) are displayed. The Rabi frequencies
$\Omega_1(t)$ and $\Omega_2(t)$ are assumed to be Gaussian envelops
for the simulations, i.e.,
$\Omega_i(t)=\Omega_ie^{-(t-\tau_i)^2/\delta\tau_i^2} (i=1,2)$.
Clearly, this process is an adiabatic passage of dark state
(\ref{DS}), since the excited state $\vert e\rangle$ of each atom
and photon states are vanishingly populated (less than $10^{-3}$).
Thus the numerical simulations clearly verify the analytical
results.

In order to quantify the robustness of this protocol against cavity
decay, atomic spontaneous emission, and fluctuations of experimental
parameters, we evaluate the succuss rate $P$ and fidelity $F$, and
compare them with those obtained in other setups
\cite{prl-92-117902,pra-72-012339}. Following the standard quantum theory of damping, we
investigate the combined influence of the cavity decay and atomic spontaneous emission on the coupled system. After tracing out the
reservoir degrees of freedom, we obtain the master equation
for the density matrix of the atom-cavity system 
\begin{eqnarray}
\dot{\rho}&=&-i[\hat{H}_I,\rho]+\kappa(2\hat{a}\rho\hat{a}^\dag-\hat{a}\hat{a}^\dag\rho-\rho\hat{a}^\dag\hat{a})\nonumber\\
&&+\sum_{i=1,2;j=0,1}\frac{\gamma}{2}(2\hat{\sigma}^i_{je}\rho\hat{\sigma}^i_{ej}-\hat{\sigma}^i_{ee}\rho-\rho\hat{\sigma}^i_{ee}).
\end{eqnarray}
The success rate $P$ is defined
as the probability of producing the entangled state $\vert
\psi\rangle=1/\sqrt{2}(\vert 01\rangle+\vert10\rangle)$, and the
fidelity $F=\langle \psi \vert \rho_a\vert \psi\rangle$, where
$\rho_a=Tr_{cavity}(\rho)$ is the final reduced density matrix of the atoms.
Fig. 3(a) shows
the succuss rate and fidelity vs. $\kappa\gamma/g^2$ for this scheme
and the setup proposed in Ref.\cite{prl-92-117902}. We see that
within the relatively strong coupling regime, the success rate is
always close to unity for this protocol, while that for the setup
proposed in Ref.\cite{prl-92-117902} is very sensitive to cavity
decay. In addition, the fidelity for the maximally entangled state
is about $99.9\%$ in our proposal, but the fidelity in
Ref.\cite{prl-92-117902} is just $93.5\%$. This is due to the fact
that the present proposal does not involve cavity field excitation
and is deterministic, requiring no measurement on the cavity field.
In Fig. 3(b), we plot the fidelity for the entangled states prepared
in the present scheme and in Ref. \cite{pra-72-012339} vs. the
fluctuation of the Rabi frequency $\delta\Omega/\Omega$. Here
$\Omega= max\{\Omega_1,\Omega_2\}$. We see that under relatively
small fluctuations of the Rabi frequency, the fidelity is still very
high for our protocol ($\geq 90\%$), but it may become very small
for the scheme in Ref. \cite{pra-72-012339}. The scheme in Ref.
\cite{pra-72-012339} relies on fractional STIRAP techniques, which
requires a precise ratio of pulse endings, thus impairing its
performance if the intensities of the two classical lasers have
small variations.

For experimental implementation of the proposal, one could utilize
the recently performed experimental setups \cite{prl-98-193601}.
There, trapped Cesium atoms could couple to a high-finesse
Fabry-Perot cavity. Cs atoms are dropped from a magneto-optical trap
into the cavity and cooled into a far off-resonant trap by an
optical lattice. The states used in the setup are ground
$\vert6S_{1/2},F=3,4\rangle$ and excited $\vert6P_{3/2},F=3'\rangle$
manifolds. The experimental parameters are $g_1\sim g_2\sim
g/(2\pi)=16$ MHz, $(\Gamma,\gamma)/2\pi=(3.8,2.6)$ MHz, and we
choose $\Delta=10g, \Omega_1\sim 100$ MHz, $\Omega_2\sim 100$ MHz,
$\tau_1\sim 3 \mu s,\tau_2\sim 1.5 \mu s$, and $\delta\tau_1\sim
1.3\mu s, \delta\tau_2\sim 1.8\mu s$. With these parameters we
obtain the fidelity up to 1 for entanglement with a total
preparation time $t\sim 2 \mu s$. The life time of the entangled
state generated by the scheme is estimated to be about $20 \mu s$.
The effective life time of the photons in the experiment is about
$1/(0.001\kappa)\sim60 \mu s$. Thus the present protocol could be
implemented with these setups. Other promising devices are
superconducting circuit devices \cite{nature-451-664}, where
superconducting qubits can be individually addressed using lasers in
the transmission-line resonators.

\section{Summary} To conclude, we have presented a protocol for
the generation of atomic entangled states in a cavity QED system. It
is based on Raman transitions or adiabatic Raman passages between
two systems to entangle the ground states of two three-state
$\Lambda$-type atoms trapped in a single mode cavity. This scheme
needs neither the measurements on cavity field nor atomic detection.
Because the atomic excited states and cavity field excitations are
never involved in this proposal, our scheme is insensitive to both
cavity decay and atomic spontaneous emission. Due to the STIRAP
techniques, this proposal is robust against fluctuations of
experimental parameters. Experimentally this protocol could be
realized with the presently available technology in cavity QED.

This work was supported by the National
Key Basic Research Program No.2010CB923102.

\newpage
\begin{figure}[h]
\centering
\includegraphics[bb=86 460 441 679,totalheight=1.7in,clip]{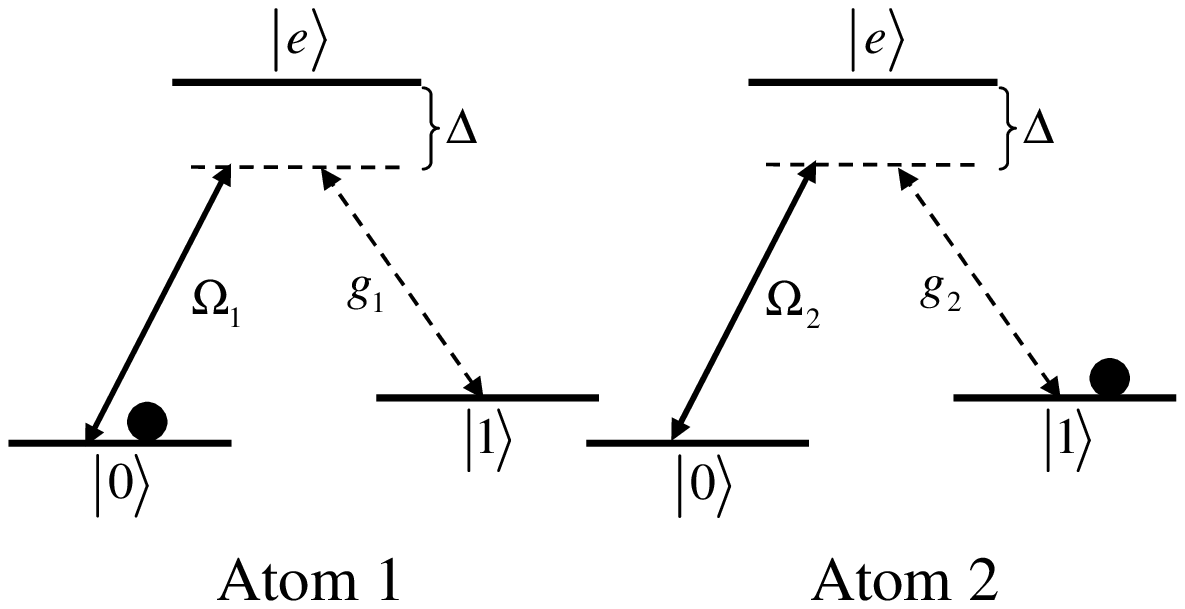}
\caption{Atomic levels of two atoms trapped in a single mode
cavity.}
\end{figure}
\newpage
\begin{figure}[h]
\centering
\includegraphics[bb=48 58 581 540,totalheight=2in,clip]{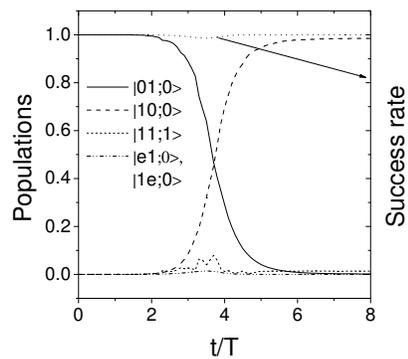}
\caption{ Populations and success rate versus time from  the full
Hamiltonian for $\Gamma=0.1g,$ and $\kappa=0.1g$, with $T=50g^{-1}$.
The parameters for the simulations are chosen as
$|\Omega_1|=2g,|\Omega_2|=g,\tau_1=0.3\times10^3g^{-1},\tau_2=0.15\times10^3g^{-1},
\delta\tau_1=0.125\times10^3g^{-1},\delta\tau_2=0.175\times10^3g^{-1}$,
and $\Delta=20g$. }
\end{figure}
\newpage
\begin{figure}[h]
\centering
\includegraphics[bb=37 57 447 740,totalheight=5.0in,clip]{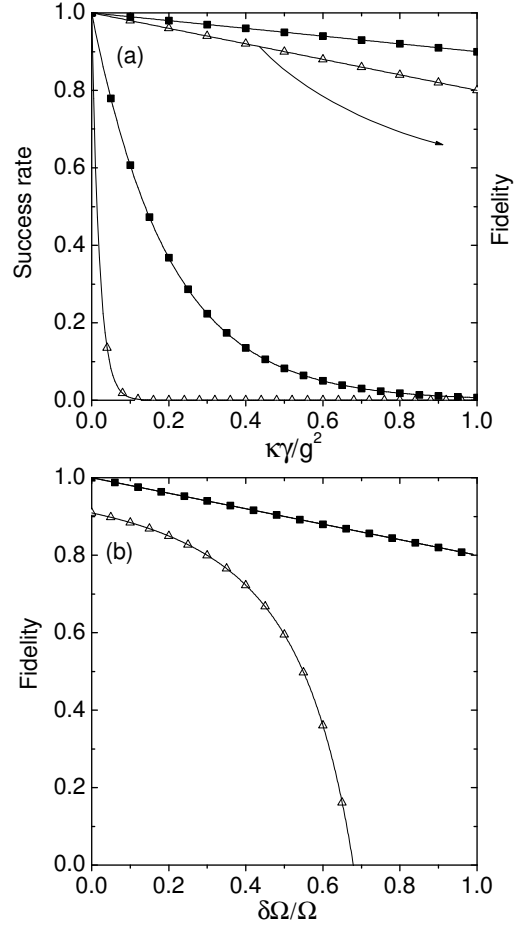}
\caption{(a) Succuss rate and fidelity vs. $\kappa\gamma/g^2$ at the
time when $|\Omega_1(t)|=|\Omega_2(t)|$. Solid square denotes the
results for this protocol, while open uptriangle corresponds to the
results for the setup in Ref.$^{17}$. (b) Fidelity vs.
$\delta\Omega/\Omega$. Solid square denotes the results for this
protocol, and open uptriangle corresponds to the results for the
setup in Ref.$^{19}$. Other parameters are chosen as in Fig. 2.}
\end{figure}

\begin{thebibliography}{23}
\expandafter\ifx\csname
natexlab\endcsname\relax\def\natexlab#1{#1}\fi
\expandafter\ifx\csname bibnamefont\endcsname\relax
  \def\bibnamefont#1{#1}\fi
\expandafter\ifx\csname bibfnamefont\endcsname\relax
  \def\bibfnamefont#1{#1}\fi
\expandafter\ifx\csname citenamefont\endcsname\relax
  \def\citenamefont#1{#1}\fi
\expandafter\ifx\csname url\endcsname\relax
  \def\url#1{\texttt{#1}}\fi
\expandafter\ifx\csname urlprefix\endcsname\relax\def\urlprefix{URL
}\fi \providecommand{\bibinfo}[2]{#2}
\providecommand{\eprint}[2][]{\url{#2}}

\bibitem{Bell}
\bibinfo{author}{\bibfnamefont{J.~S.} \bibnamefont{Bell}},
  \bibinfo{journal}{\emph{Physics} (Long Island City, NY)}
  \textbf{\bibinfo{volume}{1}}, \bibinfo{pages}{195} (1965).

\bibitem{quantum_information}
\bibinfo{author}{\bibfnamefont{M.~A.} \bibnamefont{Nielsen}} \bibnamefont{and}
  \bibinfo{author}{\bibfnamefont{I.~L.} \bibnamefont{Chuang}},
  \emph{\bibinfo{title}{Quantum Computation and Quantum Information}}
  (\bibinfo{publisher}{Cambridge University Press, Cambridge, UK},
  \bibinfo{year}{2000}).

\bibitem{Shor}
\bibinfo{author}{\bibfnamefont{P.~W.} \bibnamefont{Shor}},
  \bibinfo{journal}{\emph{SIAM J. Comput.}}
  \textbf{\bibinfo{volume}{26}},
  \bibinfo{pages}{1484} (\bibinfo{year}{1997}).

\bibitem{Prl-79-325}
\bibinfo{author}{\bibfnamefont{L.}~\bibnamefont{K.Grover}},
  \bibinfo{journal}{\emph{Phys.\ Rev. Lett.}}
  \textbf{\bibinfo{volume}{79}},
  \bibinfo{pages}{325}  (\bibinfo{year}{1997}).

\bibitem{Prl-70-1895}
\bibinfo{author}{\bibfnamefont{C.~H.} \bibnamefont{Bennett}},
  \bibinfo{author}{\bibfnamefont{G.}~\bibnamefont{Brassard}},
  \bibinfo{author}{\bibfnamefont{C.}~\bibnamefont{Crepeau}},
  \bibinfo{author}{\bibfnamefont{R.}~\bibnamefont{Jozsa}},
  \bibinfo{author}{\bibfnamefont{A.}~\bibnamefont{Peres}}, \bibnamefont{and}
  \bibinfo{author}{\bibfnamefont{W.~K.} \bibnamefont{Wootters}},
  \bibinfo{journal}{\emph{Phys.\ Rev. Lett.}}
  \textbf{\bibinfo{volume}{70}},
  \bibinfo{pages}{1895} (\bibinfo{year}{1993}).

\bibitem{Prl-67-661}
\bibinfo{author}{\bibfnamefont{A.~K.} \bibnamefont{Ekert}},
  \bibinfo{journal}{\emph{Phys.\ Rev. Lett.}} \textbf{\bibinfo{volume}{67}},
  \bibinfo{pages}{661} (\bibinfo{year}{1991}).

\bibitem{Kimble}
\bibinfo{author}{\bibfnamefont{H.~J.} \bibnamefont{Kimble}},
  \bibinfo{journal}{\emph{Phys.\ Scr.}}
  \textbf{\bibinfo{volume}{T76}},
  \bibinfo{pages}{127} (\bibinfo{year}{1998}).

\bibitem{Sci298}
\bibinfo{author}{\bibfnamefont{H.}~\bibnamefont{Mabuchi}} \bibnamefont{and}
  \bibinfo{author}{\bibfnamefont{A.~C.} \bibnamefont{Doherty}},
  \bibinfo{journal}{\emph{Science}} \textbf{\bibinfo{volume}{298}},
  \bibinfo{pages}{1372} (\bibinfo{year}{2002}).

\bibitem{mplb} Y.-C. Wang and J.-M. Liu, \emph{Int. J. Mod.\ Phys. B} \textbf{21},
2805 (2007).

\bibitem{Li} P. B Li, Y Gu, Q. H. Gong, and G. C. Guo, \emph{Phys. Rev. A} \textbf{79},
042339 (2009).

\bibitem{Guo}
\bibinfo{author}{\bibfnamefont{S.-B.} \bibnamefont{Zheng}} \bibnamefont{and}
  \bibinfo{author}{\bibfnamefont{G.-C.} \bibnamefont{Guo}},
  \bibinfo{journal}{\emph{Phys.\ Rev. Lett}.}
  \textbf{\bibinfo{volume}{85}},
  \bibinfo{pages}{2392} (\bibinfo{year}{2000}).

\bibitem{prl-87-037902}
\bibinfo{author}{\bibfnamefont{S.}~\bibnamefont{Osnaghi}},
  \bibinfo{author}{\bibfnamefont{P.}~\bibnamefont{Bertet}},
  \bibinfo{author}{\bibfnamefont{A.}~\bibnamefont{Auffeves}},
  \bibinfo{author}{\bibfnamefont{P.}~\bibnamefont{Maioli}},
  \bibinfo{author}{\bibfnamefont{M.}~\bibnamefont{Brune}},
  \bibinfo{author}{\bibfnamefont{J.~M.} \bibnamefont{Raimond}},
  \bibnamefont{and} \bibinfo{author}{\bibfnamefont{S.}~\bibnamefont{Haroche}},
  \bibinfo{journal}{\emph{Phys.\ Rev. Lett.}}
  \textbf{\bibinfo{volume}{87}},
  \bibinfo{pages}{037902} (\bibinfo{year}{2001}).

\bibitem{Prl-89-237901}
\bibinfo{author}{\bibfnamefont{J.}~\bibnamefont{Hong}} \bibnamefont{and}
  \bibinfo{author}{\bibfnamefont{H.-W.} \bibnamefont{Lee}},
  \bibinfo{journal}{\emph{Phys.\ Rev. Lett}.}
  \textbf{\bibinfo{volume}{89}},
  \bibinfo{pages}{237901} (\bibinfo{year}{2002}).

\bibitem{Prl-90-127903} M. Yang, Y.-M. Yi, and Z.-L. Cao, \emph{Int. J. Quantum.
Inf.} \textbf{2}, 231 (2004).


\bibitem{Prl-90-217902}
\bibinfo{author}{\bibfnamefont{X.-L.} \bibnamefont{Feng}},
  \bibinfo{author}{\bibfnamefont{Z.-M.} \bibnamefont{Zhang}},
  \bibinfo{author}{\bibfnamefont{X.-D.} \bibnamefont{Li}},
  \bibinfo{author}{\bibfnamefont{S.-Q.} \bibnamefont{Gong}}, \bibnamefont{and}
  \bibinfo{author}{\bibfnamefont{Z.-Z.} \bibnamefont{Xu}},
  \bibinfo{journal}{\emph{Phys.\ Rev. Lett.}}
  \textbf{\bibinfo{volume}{90}},
  \bibinfo{pages}{217902} (\bibinfo{year}{2003}).

\bibitem{Prl-90-253601}
\bibinfo{author}{\bibfnamefont{L.-M.} \bibnamefont{Duan}} \bibnamefont{and}
  \bibinfo{author}{\bibfnamefont{H.~J.} \bibnamefont{Kimble}},
  \bibinfo{journal}{\emph{Phys.\ Rev. Lett.}}
  \textbf{\bibinfo{volume}{90}},
  \bibinfo{pages}{253601} (\bibinfo{year}{2003}).

\bibitem{Prl-91-067901}
\bibinfo{author}{\bibfnamefont{D.~E.} \bibnamefont{Browne}},
  \bibinfo{author}{\bibfnamefont{M.~B.} \bibnamefont{Plenio}},
  \bibnamefont{and} \bibinfo{author}{\bibfnamefont{S.~F.}
  \bibnamefont{Huelga}}, \bibinfo{journal}{\emph{Phys.\ Rev. Lett}.}
  \textbf{\bibinfo{volume}{91}} \bibinfo{pages}{067901} (\bibinfo{year}{2003}).

\bibitem{prl-92-117902}
\bibinfo{author}{\bibfnamefont{C.-P.} \bibnamefont{Yang}},
  \bibinfo{author}{\bibfnamefont{S.-I.} \bibnamefont{Chu}}, \bibnamefont{and}
  \bibinfo{author}{\bibfnamefont{S.}~\bibnamefont{Han}},
  \bibinfo{journal}{\emph{Phys.\ Rev. Lett.}}
  \textbf{\bibinfo{volume}{92}}
  \bibinfo{pages}{117902} (\bibinfo{year}{2004}).

\bibitem{pra-68-033817}
\bibinfo{author}{\bibfnamefont{C.}~\bibnamefont{Marr}},
  \bibinfo{author}{\bibfnamefont{A.}~\bibnamefont{Beige}}, \bibnamefont{and}
  \bibinfo{author}{\bibfnamefont{G.}~\bibnamefont{Rempe}},
  \bibinfo{journal}{\emph{Phys.\ Rev. A}}
  \textbf{\bibinfo{volume}{68}},
  \bibinfo{pages}{033817} (\bibinfo{year}{2003}).

\bibitem{pra-72-012339}
\bibinfo{author}{\bibfnamefont{M.}~\bibnamefont{Amniat-Talab}},
  \bibinfo{author}{\bibfnamefont{S.}~\bibnamefont{Gu\'{e}rin}},
  \bibnamefont{and} \bibinfo{author}{\bibfnamefont{H.~R.}
  \bibnamefont{Jauslin}}, \bibinfo{journal}{\emph{Phys.\ Rev. A}}
  \textbf{\bibinfo{volume}{72}}, \bibinfo{pages}{012339} (\bibinfo{year}{2005}).

\bibitem{RMP-70-1003}
\bibinfo{author}{\bibfnamefont{K.}~\bibnamefont{Bergmann}},
  \bibinfo{author}{\bibfnamefont{H.}~\bibnamefont{Theuer}}, \bibnamefont{and}
  \bibinfo{author}{\bibfnamefont{B.~W.} \bibnamefont{Shore}},
  \bibinfo{journal}{\emph{Rev.\ Mod.\ Phys}.}
  \textbf{\bibinfo{volume}{70}},
  \bibinfo{pages}{1003} (\bibinfo{year}{1998}).

\bibitem{James}
\bibinfo{author}{\bibfnamefont{D.~F.~V.} \bibnamefont{James}},
  \bibinfo{journal}{\emph{Fortschr. Phys.}}
  \textbf{\bibinfo{volume}{48}},
  \bibinfo{pages}{823} (\bibinfo{year}{2000}).

\bibitem{prl-98-193601}
\bibinfo{author}{\bibfnamefont{A.~D.} \bibnamefont{Boozer}},
  \bibinfo{author}{\bibfnamefont{A.}~\bibnamefont{Boca}},
  \bibinfo{author}{\bibfnamefont{R.}~\bibnamefont{Miller}},
  \bibinfo{author}{\bibfnamefont{T.~E.} \bibnamefont{Northup}},
  \bibnamefont{and} \bibinfo{author}{\bibfnamefont{H.~J.}
  \bibnamefont{Kimble}}, \bibinfo{journal}{\emph{Phys.\ Rev. Lett.}}
  \textbf{\bibinfo{volume}{98}}, \bibinfo{pages}{193601} (\bibinfo{year}{2007}).

\bibitem{nature-451-664}
\bibinfo{author}{\bibfnamefont{R.~J.} \bibnamefont{Schoelkopf}}
  \bibnamefont{and} \bibinfo{author}{\bibfnamefont{S.~M.}
  \bibnamefont{Girvin}}, \bibinfo{journal}{\emph{Nature }(London)}
  \textbf{\bibinfo{volume}{451}}, \bibinfo{pages}{664} (\bibinfo{year}{2008}).

\end{thebibliography}
\end{document}